\documentclass[floats,floatfix,showpacs,amssymb,prl,twocolumn,superscriptaddress,nofootinbib]{revtex4-1}

\newlength{\figw} 
\usepackage{graphicx,epsf, epsfig, amssymb}
\usepackage{bm}
\usepackage{longtable,tabularx}
\usepackage{color}
\usepackage[breaklinks]{hyperref}
\usepackage{amsfonts,amsmath,amssymb,mathrsfs,gensymb}
\usepackage{natbib}
\usepackage{aas_macros}
\usepackage{array}
\usepackage{multirow}
\usepackage{rotating,array}
\usepackage{graphicx}

\newcommand\optional[1]{}
\def\be{\begin{equation}}
\def\ee{\end{equation}}
\def\beq{\begin{eqnarray}}
\def\eeq{\end{eqnarray}}
\usepackage{comment}

\begin{document}

\title{Effective Potentials and Morphological Transitions for Binary Black Hole Spin Precession}

\author{Michael Kesden}
\email{kesden@utdallas.edu}
\affiliation{Department of Physics, The University of Texas at Dallas, Richardson, TX 75080, USA }

\author{Davide Gerosa}
\email{d.gerosa@damtp.cam.ac.uk}
\affiliation{Department of Applied Mathematics and Theoretical Physics, Centre for Mathematical Sciences, University of Cambridge, Wilberforce Road, Cambridge CB3 0WA, UK}

\author{Richard O'Shaughnessy}
\email{rossma@rit.edu}
\affiliation{Center for Computational Relativity and Gravitation, Rochester Institute of Technology, Rochester, NY 14623, USA}

\author{Emanuele Berti}
\email{eberti@olemiss.edu}
\affiliation{Department of Physics and Astronomy, The University of
Mississippi, University, MS 38677, USA}

\author{Ulrich Sperhake}
\email{U.Sperhake@damtp.cam.ac.uk}
\affiliation{Department of Applied Mathematics and Theoretical Physics, Centre for Mathematical Sciences, University of Cambridge, Wilberforce Road, Cambridge CB3 0WA, UK}
\affiliation{Department of Physics and Astronomy, The University of
Mississippi, University, MS 38677, USA}
\affiliation{California Institute of Technology, Pasadena, CA 91125, USA}


\date{\today}

\begin{abstract}

We derive an effective potential for binary black-hole (BBH) spin precession at second post-Newtonian order.  This effective
potential allows us to solve the orbit-averaged spin-precession equations analytically for arbitrary mass ratios and spins. 
These solutions are quasiperiodic functions of time: after a fixed period the BBH spins return to their initial relative
orientations and jointly precess about the total angular momentum by a fixed angle.  Using these solutions, we classify BBH
spin precession into three distinct morphologies between which BBHs can transition during their inspiral.  We also derive a
precession-averaged evolution equation for the total angular momentum that can be integrated on the radiation-reaction
time and identify a new class of spin-orbit resonances that can tilt the direction of the total angular momentum during the
inspiral.  Our new results will help efforts to model and interpret gravitational waves from generic BBH mergers and predict
the distributions of final spins and gravitational recoils.  

\end{abstract}
\maketitle 

\noindent{\em Introduction.~--~}The classic two-body problem was a major engine of historical progress in physics and
astronomy.  This problem can be solved analytically in Newtonian gravity; its solutions are the well known Keplerian orbits.
The analogs to Newtonian point masses in general relativity are binary black holes (BBHs).  Astrophysical BBHs have
spins $\mathbf{S}_i$ \cite{1963PhRvL..11..237K} in addition to their masses $m_i$ [the masses determine the total mass
$M \equiv m_1 + m_2$, mass ratio $q \equiv m_2/m_1 \leq 1$, and symmetric mass ratio $\eta \equiv m_1m_2/M^2 =
q/(1+q)^2$].  Full solutions to the two-body problem in general relativity must therefore include spin evolution in addition to
orbital motion.  Einstein's equations must be
solved numerically \cite{2005PhRvL..95l1101P,2006PhRvL..96k1101C,2006PhRvL..96k1102B} when the binary
separation $r$ is comparable to the gravitational radius $r_g \equiv GM/c^2$, but post-Newtonian (PN) approximations
may be used when $r \gg r_g$.  BBH evolution in the PN limit occurs on three distinct timescales: the orbital time
$t_{\rm orb} \sim (r^3/GM)^{1/2}$ on which the binary separation $\mathbf{r}$ evolves, the precession time
$t_{\rm pre} \sim c^2r^{5/2}/[\eta(GM)^{3/2}] \sim (t_{\rm orb}/\eta)(r/r_g)$ on which the spin directions change, and the
radiation-reaction time $t_{\rm RR} \sim E/|dE_{\rm GW}/dt| \sim c^5 r^4/[\eta(GM)^3] \sim (t_{\rm orb}/\eta)(r/r_g)^{5/2}$ on
which the energy $E=-G\eta M^2/(2r)$ and orbital angular momentum $L=\eta (r G M^3)^{1/2}$ decrease.

The hierarchy $t_{\rm orb} \ll t_{\rm pre} \ll t_{\rm RR}$ implies that when considering evolution on one timescale,
quantities evolving on a shorter (longer) timescale can be averaged (held constant).  This has been used to derive
orbit-averaged spin-precession equations $\dot{\mathbf{S}}_i = \Bar{\boldsymbol{\Omega}}_i \times \mathbf{S}_i$
\cite{1979GReGr..11..149B,1985PhRvD..31.1815T,1994PhRvD..49.6274A,1995PhRvD..52..821K},
where the precession frequencies $\Bar{\boldsymbol{\Omega}}_i$ depend on the orbital angular momentum $\mathbf{L}$
and spins $\mathbf{S}_i$ but not
on the instantaneous separation $\mathbf{r}$.  These equations can be integrated numerically with time steps $t_{\rm orb}
\ll \Delta t \lesssim t_{\rm pre}$, greatly reducing the computational cost of evolving spin directions for many orbital
times.  In this {\it Letter}, we show that the 2PN spin-precession equations \cite{2008PhRvD..78d4021R}
can be solved {\it analytically} in a suitably chosen
frame for arbitrary mass ratios $q$ and spins $\mathbf{S}_i$.  The relative orientations of $\mathbf{L}$
and $\mathbf{S}_i$ are fully specified by the angles $\theta_i$ between $\mathbf{L}$ and $\mathbf{S}_i$ and the angle
$\Delta\Phi$ between the spin components in the orbital plane; we provide parametric solutions for these angles in terms
of $S$, the magnitude of the total spin $\mathbf{S} = \mathbf{S}_1 + \mathbf{S}_2$.  These solutions improve our
understanding of spin precession in much the same way that the solutions $r(f) = a(1-e^2)/(1+e\cos f)$ for Keplerian
orbits provide additional insight beyond Newton's law $\ddot{\mathbf{r}} = -GM\hat{\mathbf{r}}/r^2$.  We can use these
solutions to {\it precession-average} the radiation-reaction equations for $dE/dt$ and $d\mathbf{J}/dt$, allowing them to be
numerically integrated with a time step $t_{\rm pre} \ll \Delta t^\prime \lesssim t_{\rm RR}$.  This greatly reduces the
computational cost of evolving BBHs compared to the previous approach that integrated the orbit-averaged precession
equations with the shorter time step $\Delta t \ll \Delta t^\prime$.  This improved efficiency is essential for transferring the
BBH spins predicted at formation by population-synthesis models \cite{2005ApJ...632.1035O,2008ApJ...682..474B,2010ApJ...719L..79F,2013PhRvD..87j4028G,2003ApJ...585L.101H,2008ApJ...684..822B,2012MNRAS.423.2533B,2014ApJ...794..104S,Sijacki:2014yfa} to near merger, where spin directions affect gravitational waves (GWs) with
frequencies in the sensitivity bands of current and future GW detectors
\cite{2010CQGra..27h4006H,2013IJMPD..2241010U,2012CQGra..29l4007S,2010CQGra..27s4002P,2013GWN.....6....4A,2014arXiv1410.2907B}.
Our new solutions may also facilitate the construction and interpretation of GW signals from BBHs in which both spins are
misaligned. In particular, stellar-mass BBH spins depend on BH natal kicks and binary stellar evolution that can thus be
constrained by ground-based GW detectors \cite{2013PhRvD..87j4028G,2014PhRvD..89l4025G}.
Hereafter we use geometrical units $G=c=1$.

\noindent{\em Precessional Solutions.~--~}Consider the evolution of BBHs with misaligned spins on a circular orbit \cite{1963PhRv..131..435P} on timescales
$t_{\rm pre} < t \ll t_{\rm RR}$.  We choose $\hat{\mathbf{z}}$ parallel to the total angular momentum $\mathbf{J}$,
$\hat{\mathbf{x}}$ parallel to the component of the orbital angular momentum $\mathbf{L}$ perpendicular to $\mathbf{J}$,
and $\hat{\mathbf{y}} = \hat{\mathbf{z}} \times \hat{\mathbf{x}}$ to complete the orthonormal triad.  Since $\mathbf{S} =
\mathbf{J} - \mathbf{L}$, it too must lie in the $xz$--plane.  Although $\mathbf{J}$ and thus $\hat{\mathbf{z}}$ are conserved
on the precession time $t_{\rm pre}$, $\hat{\mathbf{x}}$ and $\hat{\mathbf{y}}$ precess about $\hat{\mathbf{z}}$ with a
frequency $\Omega_z$.  The angle $\theta_L$ between $\mathbf{J}$ and $\mathbf{L}$ is given by
\begin{equation} \label{E:csthL}
\cos\theta_L = \frac{J^2 + L^2 - S^2}{2JL}
\end{equation}
and depends exclusively on $S$ and the constants $L$ and $J$.  We define a second orthonormal frame such that
$\hat{\mathbf{z}}^\prime = \hat{\mathbf{S}}$, $\hat{\mathbf{y}}^\prime = \hat{\mathbf{y}}$, and $\hat{\mathbf{x}}^\prime =
\hat{\mathbf{y}}^\prime \times \hat{\mathbf{z}}^\prime$ completes the triad.  $\mathbf{S}_1$ points in the direction
$(\theta^\prime, \varphi^\prime)$ specified by traditional spherical coordinates in this second frame, where
\begin{equation} \label{E:csthpr}
\cos\theta^\prime = \frac{S^2 + S_1^2 - S_2^2}{2SS_1}
\end{equation}
also depends only on $S$, because $S_1$ and $S_2$ are conserved.  Since $\mathbf{S}_2 = \mathbf{S} - \mathbf{S}_1$,
the directions of all these angular momenta are specified in our initial (unprimed) frame by $S$ and $\varphi^\prime$.

The projected effective spin \cite{2001PhRvD..64l4013D}
\begin{equation} \label{E:xidef}
\xi \equiv M^{-2} [(1+q)\mathbf{S}_1 + (1+q^{-1})\mathbf{S}_2] \cdot \hat{\mathbf{L}}
\end{equation}
is conserved by 2PN spin precession \cite{2008PhRvD..78d4021R}. Radiation reaction (at 2.5PN) preserves the direction of
$\mathbf{L}$ and thus $\xi$ is further conserved on the radiation-reaction time $t_{\rm RR}$ as seen in our
previous work \cite{2010PhRvD..81h4054K}.  Inserting expressions for $\mathbf{L}$ and $\mathbf{S}_i$ in terms of $S$ and
$\varphi^\prime$ into Eq.~(\ref{E:xidef}) yields
\begin{align} \label{E:xiSP}
\xi(S, \varphi^\prime) &= \{ (J^2 - L^2 - S^2)[S^2(1+q)^2 - (S_1^2 - S_2^2)(1- q^2)] \notag \\
& \quad - (1- q^2) A_1A_2A_3A_4\cos\varphi^\prime \}/(4qM^2S^2L)\,,
\end{align}
where
\begin{subequations}
\begin{align}
A_1 &\equiv [J^2 - (L - S)^2]^{1/2}\,, \\
A_2 &\equiv [(L + S)^2 - J^2]^{1/2}\,, \\
A_3 &\equiv [S^2 - (S_1 - S_2)^2]^{1/2}\,, \\
A_4 &\equiv [(S_1 + S_2)^2 - S^2]^{1/2}\,.
\end{align}
\end{subequations}
The $A_i$'s are real in the allowed range $J_{\rm min} \leq J \leq J_{\rm max}$, $S_{\rm min} \leq S \leq S_{\rm max}$,
where
\begin{subequations}
\begin{align}
J_{\rm min} &= L - S_1 - S_2\,, \\
J_{\rm max} &= L + S_1 + S_2\,, \\
S_{\rm min} &= {\rm max} \{ |J - L|, |S_1 - S_2| \}\,, \\
S_{\rm max} &= {\rm min} \{ J + L, S_1 + S_2 \}\,.
\end{align}
\end{subequations}
Our approach needs to be modified for $L \leq S_1 + S_2$  ($J_{\rm min} \leq 0$ above)
\cite{1994PhRvD..49.6274A,GerosaPrep}, but this does not occur until $r \leq r_{\rm min} = [(1+q^2)/q]^2 r_g$ for maximally
spinning BBHs ($r_{\rm min} = 4r_g$ for $q = 1$).  Eq.~(\ref{E:xiSP}) can be solved for $\cos\varphi^\prime$ and then
inserted into expressions for $\mathbf{L}$ and $\mathbf{S}_i$ to obtain the surprisingly simple relations
\begin{subequations} \label{E:3sol}
\begin{align}
\cos\theta_1 &=  \frac{1}{2(1-q)S_1} \left[ \frac{J^2 - L^2 -S^2}{L} - \frac{2qM^2\xi}{1+q} \right]\,, \label{E:cs1} \\
\cos\theta_2 &=  \frac{q}{2(1-q)S_2} \left[ -\frac{J^2 - L^2 -S^2}{L} + \frac{2M^2\xi}{1+q} \right]\,, \label{E:cs2} \\
\cos\theta_{12} &= \frac{S^2 - S_1^2 - S_2^2}{2S_1S_2}\,, \\
\cos\Delta\Phi &= \frac{\cos\theta_{12} - \cos\theta_1\cos\theta_2}{\sin\theta_1\sin\theta_2} 
\end{align}
\end{subequations}
for $\cos\theta_i \equiv \hat{\mathbf{L}} \cdot \hat{\mathbf{S}}_i$, $\cos\theta_{12} \equiv \hat{\mathbf{S}}_1\cdot
\hat{\mathbf{S}}_2$,  and the angle $\Delta\Phi$ between the spin components in the orbital plane.  Note that $S$ is the only
variable in these expressions that evolves on $t_{\rm pre}$.


\begin{figure}[thb]
\includegraphics[width=0.5\textwidth]{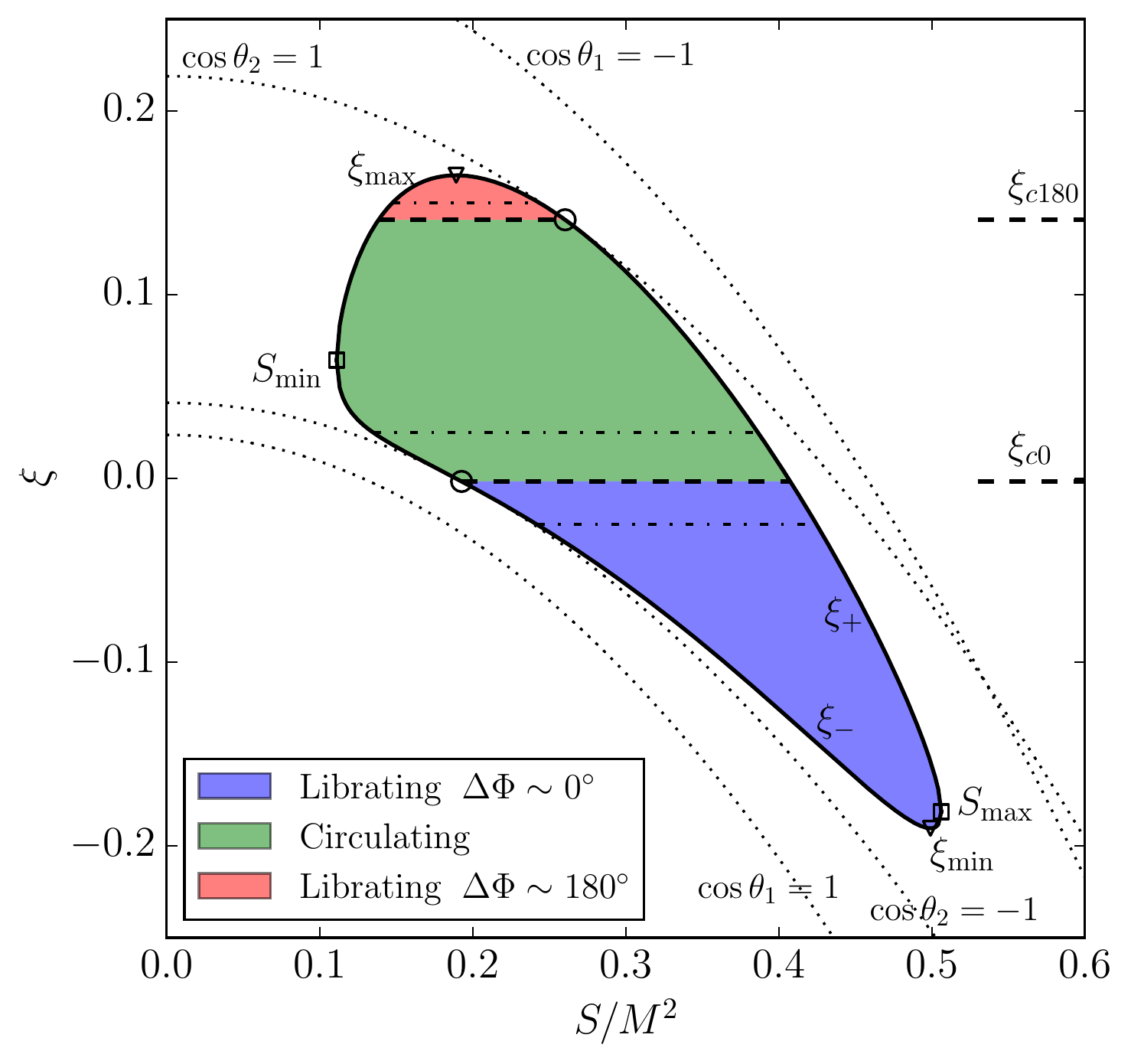}
\caption{Effective potentials $\xi_\pm(S)$ for the spin precession of BBHs with maximal spins, mass ratio
$q = 0.8$, $L = 0.781 M^2$ ($r = 10M$), and $J = 0.85M^2$.  These two functions form a loop enclosing the allowed values
of $S$ and $\xi$.  Since $\xi$ is conserved during the inspiral, $S$ oscillates between the two roots $S_\pm$ of the equation
$\xi = \xi_\pm(S)$ on the precession time.  The two roots are degenerate at $\xi_{\rm min}$ and $\xi_{\rm max}$, implying
that $S$ is constant: these configurations correspond respectively to the $\Delta\Phi = 0^\circ (\Delta\Phi=\pm180^\circ)$
spin-orbit resonances of Schnittman \cite{2004PhRvD..70l4020S}.  The four dotted curves are the contours $\cos\theta_i =
\pm1$ given by Eqs.~(\ref{E:cs1}) and (\ref{E:cs2}); transitions between BBHs for which $\Delta\Phi$ circulates and those for
which it librates about $0^\circ$ ($\pm180^\circ$) occur where these curves are tangent to the potentials $\xi_\pm(S)$, as
indicated by the lower (upper) dashed line $\xi = \xi_{c0}$ ($\xi_{c180}$).  The three dot-dashed lines correspond to the
three BBH systems shown in Fig.~\ref{F:arrows} as representative of each morphology.}
\label{F:Sxi}
\end{figure}

The evolution of $S$ is also surprisingly simple.  If we set $\cos\varphi^\prime = \mp1$ in Eq.~(\ref{E:xiSP}), we obtain two
functions $\xi_\pm(S)$ that act like effective potentials for $S$.  For given values of $L$ and $J$, one of the $A_i$'s
vanishes at $S_{\rm min}$ and $S_{\rm max}$, implying that $\xi_+(S_{\rm min}) = \xi_-(S_{\rm min})$, $\xi_+(S_{\rm max})
= \xi_-(S_{\rm max})$. Thus the two curves $\xi_\pm(S)$ form a closed loop in the $S\xi$--plane, as shown in
Fig~\ref{F:Sxi}.  The equation $\xi = \xi_\pm(S)$ has two roots $S_\pm(L, J, \xi)$ that determine the allowed range $S_- \leq
S \leq S_+$.  This is entirely analogous to how two roots $r_\pm(E, L)$ of the equation $E = V(r, L)$, where $V$ is the
effective potential for radial motion, determine pericenter and apocenter.  The two roots are degenerate ($S_- = S_+$) at the
maximum $\xi_{\rm max}(L, J)$ of $\xi_+(S)$ and minimum $\xi_{\rm min}(L, J)$ of $\xi_-(S)$, implying that $S = S_\pm$
remains constant -- just as $r$ remains constant for values of $E$ and $L$ corresponding to circular orbits, the minimum of
the effective potential $V(r, L)$.  These two configurations ($\xi = \xi_{\rm min}$ and $\xi = \xi_{\rm max}$) are precisely the
$\Delta\Phi = 0^\circ$ and $\Delta\Phi = \pm180^\circ$ spin-orbit resonances identified by Schnittman
\cite{2004PhRvD..70l4020S}.

\begin{figure*}[thb]
\includegraphics[width=0.8\textwidth]{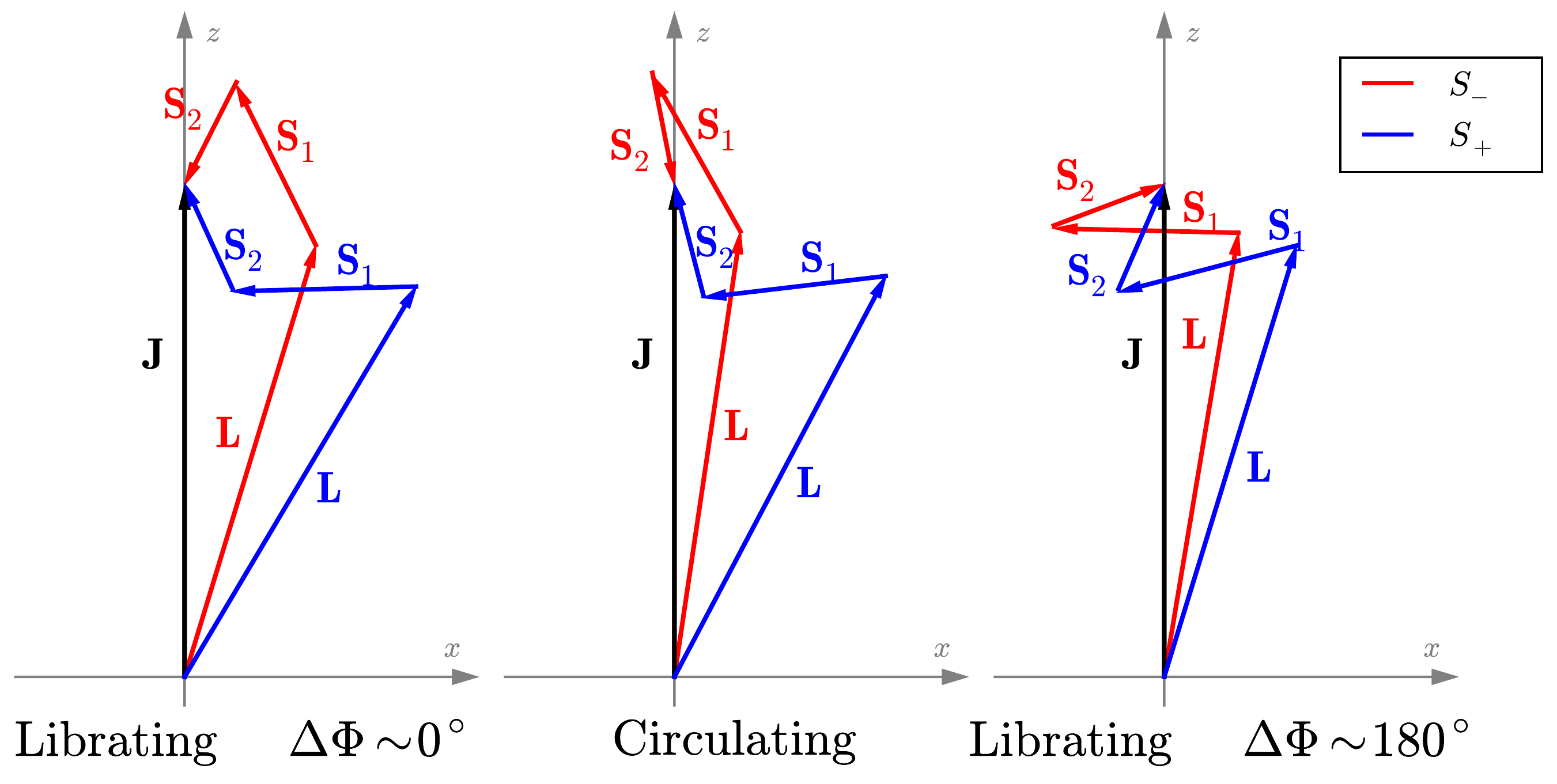}
\caption{The three morphologies of BBH spin precession.  The angular momenta $\mathbf{J}$, $\mathbf{L}$, and
$\mathbf{S}_i$ are all in the $xz$--plane at $S = S_\pm$. In all three panels the BBHs have maximal spins, $q = 0.8$,
$L = 0.781 M^2$ ($r = 10M$), and $J = 0.85M^2$ as in Fig.~\ref{F:Sxi}. The left, middle, and right panels correspond to
$\xi = -0.025$, $0.025$ and $0.15$,
respectively.  If the components of $\mathbf{S}_i$ perpendicular to $\mathbf{L}$ are aligned with each other at both roots
$S_\pm$, $\Delta\Phi$ librates about $0^\circ$.  If they are aligned at one root and anti-aligned at the other, $\Delta\Phi$
circulates.  If they are anti-aligned at both roots, $\Delta\Phi$ librates about $180^\circ$.}
\label{F:arrows}
\end{figure*}

The BBH spins $\mathbf{S}_i$ and orbital angular momentum $\mathbf{L}$ are shown at $S = S_\pm$ for three
different values of $\xi$ but the same $L$ and $J$ in Fig.~\ref{F:arrows}.  These vectors are coplanar at $S_\pm$, since these
points lie on the curves $\xi_\pm(S)$ defined such that $\cos\varphi^\prime = \mp1$; we must therefore have
$\Delta\Phi = 0^\circ$ or $180^\circ$ at $S_\pm$.  There are three possibilities as $S$ increases from $S_-$ to $S_+$
and returns to $S_-$: (1) $\Delta\Phi$ begins at $0^\circ$, decreases to a minimum $-\Delta\Phi_\ast$, returns to $0^\circ$ at
$S_+$, increases to a maximum $+\Delta\Phi_\ast$, then returns to $0^\circ$ back at $S_-$, (2) $\Delta\Phi$ begins at
$-180^\circ$, increases to $0^\circ$ at $S_+$, then continues to increase to $+180^\circ$ back at $S_-$, and (3)
$\Delta\Phi$ begins at $180^\circ$, increases to a maximum $180^\circ+\Delta\Phi_\ast$, returns to $180^\circ$ at $S_+$,
decreases to a minimum $180^\circ-\Delta\Phi_\ast$, then returns to $180^\circ$ back at $S_-$.  These three possibilities
(libration about $\Delta\Phi = 0^\circ$, circulation, and libration about $\Delta\Phi = 180^\circ$) are shown in the left, center,
and right panels of Fig.~\ref{F:arrows}.  The libration amplitude $\Delta\Phi_\ast$ depends on $L$, $J$, and $\xi$.

Eq.~(\ref{E:3sol}) implies that BBHs with $\xi = \xi_{\rm min}$ are trapped in the $\Delta\Phi = 0^\circ$ resonance.
Comparing the left and center panels of Fig.~\ref{F:arrows}, we see that the transition between BBHs with $\Delta\Phi =
0^\circ$ and those with $\Delta\Phi = \pm180^\circ$ at $S_-$ [$(1) \to (2)$ above] occurs at the value $\xi \equiv \xi_{c0}$
at which $\mathbf{L}$ is aligned with either $\mathbf{S}_1$ or $-\mathbf{S}_2$ at $S_-$.  This transition is marked by the
lower dashed line separating the blue and green regions in Fig.~\ref{F:Sxi}.  As $\xi$ increases further, we see by comparing
the center and right panels of Fig.~\ref{F:arrows} that we eventually reach a value $\xi \equiv \xi_{c180}$ at which
$\Delta\Phi$ transitions from $0^\circ$ to $180^\circ$ at $S_+$ [$(2) \to (3)$ above].  This transition occurs when
$\mathbf{L}$ is aligned with either $\mathbf{S}_2$ or $-\mathbf{S}_1$ at $S_+$ and is marked by the upper dashed line
separating the green and red regions in Fig.~\ref{F:Sxi}.  These morphological transitions correspond to the quasistable
equilibria noted by Schnittman \cite{2004PhRvD..70l4020S}.  Finally, as $\xi$ continues to increase the amplitude of the
oscillations in $S$ decreases, until the $\Delta\Phi = \pm180^\circ$ resonance is reached at $\xi_{\rm max}$.

Although $S$ parameterizes spin directions much like the true anomaly parameterizes Keplerian orbits, one may also
want the time-dependent solutions $S(t)$.  The spin-precession equations
\cite{1995PhRvD..52..821K,2008PhRvD..78d4021R,2006PhRvD..74j4033F,2006PhRvD..74j4034B} imply
\begin{align} \label{E:dSdt}
\frac{dS}{dt} &= -\frac{3(1-q^2)}{2q} \frac{S_1S_2}{S} \frac{(\eta^2M^3)^3}{L^5} \left( 1 - \frac{\eta M^2 \xi}{L}  \right)
\notag \\
& \times \sin\theta_1 \sin\theta_2 \sin\Delta\Phi
\end{align}
where again the right-hand side depends only on $S$ when we use Eq.~(\ref{E:3sol}).  Oscillations in $S$ have a precessional period
$\tau(L, J, \xi) = 2\int_{S_-}^{S_+}dS/|dS/dt|$.  The basis vectors $\hat{\mathbf{x}}$ and $\hat{\mathbf{y}}$ precess about
$\hat{\mathbf{z}}$ at a rate
\begin{align} \label{E:Omegaz}
\Omega_z &= \frac{J}{2} \left( \frac{\eta^2M^3}{L^2} \right)^3 \bigg\{ 1 + \frac{3}{2\eta} \left( 1 - \frac{\eta M^2 \xi}{L} \right)
\notag \\ 
& -\frac{3(1+q)}{2qA_1^2A_2^2} \left(1 - \frac{\eta M^2 \xi}{L} \right)[4(1-q)L^2(S_1^2 - S_2^2)
\notag \\
& -(1+q)(J^2 - L^2 -S^2)(J^2 - L^2 -S^2 - 4\eta M^2L\xi)] \bigg\}\,,
\end{align}
implying that they precess through an angle $\alpha(L, J, \xi) = 2\int_{S_-}^{S_+}(\Omega_z~dS)/|dS/dt|$ in each
precessional period.

\noindent{\em Gravitational Inspiral.~--~}Although $L$ and $J$ are conserved on $t_{\rm pre}$, they vary on the longer radiation-reaction timescale $t_{\rm RR}$.
At lowest PN order, the orbit-averaged angular momentum flux is given by the well known quadrupole formula
\cite{1963PhRv..131..435P}  $d\mathbf{J}/dt = -(32/5)(\eta M^2/L)^8 (\eta\mathbf{L}/M)$, implying $dL/dt =
\hat{\mathbf{L}} \cdot d\mathbf{J}/dt$ and $dJ/dt = \hat{\mathbf{J}} \cdot d\mathbf{J}/dt$.  This expression for $dL/dt$ is
independent of $S$, but that for $dJ/dt$ is not.  However, if the above precession angle $\alpha \neq 2\pi n$ for integer
$n$, the average of $d\mathbf{J}/dt$ over many precession periods will be parallel to $\mathbf{J}$.  Using the
monotonically decreasing $L$ to parameterize the inspiral we obtain the {\it precession-averaged} result
\begin{align}
\left\langle \frac{dJ}{dL} \right\rangle_{\rm pre} &= \frac{2}{\tau} \int_{S_-}^{S_+} \frac{\cos\theta_L ~dS}{|dS/dt|} \notag \\
&= \frac{1}{2LJ} \left[ J^2 + L^2 - \frac{2}{\tau} \int_{S_-}^{S_+} \frac{S^2~dS}{|dS/dt|} \right]\,,
\end{align}
that is indeed independent of $S$. At higher PN order, $d\mathbf{J}/dt$ is spin-dependent and thus $\langle dJ/dL
\rangle_{\rm pre}$ is not simply a time-weighted average of $\cos\theta_L$ \cite{1995PhRvD..52..821K}.
This equation allows $J$ to be evolved numerically with a time step $t_{\rm pre} \ll
\Delta t^\prime \lesssim t_{\rm RR}$ consistent with the timescale on which it is varying.  Unless we need to keep track
of the precessional phase, we can use this precession-averaged equation combined with the orbit-averaged solutions of
the previous section to evolve BBH spin directions far more efficiently than the conventional approach relying
exclusively on the orbit-averaged equations $\dot{\mathbf{S}}_i = \Bar{\boldsymbol{\Omega}}_i \times \mathbf{S}_i$.
Preliminary results \cite{GerosaPrep} indicate that as $L$ and $J$ evolve, circulating BBHs
[$\xi_{\rm c0}(L, J) \leq \xi \leq \xi_{\rm c180}(L, J)$] can be captured into one of the two librating morphologies
[$\xi < \xi_{\rm c0}(L, J)$ or $\xi > \xi_{\rm c180}(L, J)$] consistent with earlier studies \cite{2004PhRvD..70l4020S}.

The condition $\alpha(L, J, \xi) =  2\pi n$ corresponds to a newly identified resonance between precession about
$\mathbf{J}$ and precession in the meridional plane (the $xz$--plane in our basis).  The direction of $\mathbf{J}$
evolves rapidly at these resonances, since $\langle d\mathbf{J}/dt \rangle_{\rm pre} \nparallel \mathbf{J}$.  This could
affect the direction of the spin of the final black hole, which is often assumed to point in the direction of $\mathbf{J}$ at
merger \cite{2010PhRvD..81h4054K,2008PhRvD..77b6004B,2009ApJ...704L..40B}, and could also leave an
observational signature if a resonance occurs within the sensitivity band of GW detectors.  Preliminary results
\cite{ZhaoPrep} suggest that generic BBHs often pass through these resonances as they inspiral.

\noindent{\em Discussion.~--~}We have derived new analytic solutions for BBH spin precession by recognizing that $L$, $J$, and $\xi$ remain constant
on the precession time $t_{\rm pre}$.  These solutions provide new insights into this deeply fundamental problem in general relativity and
allow us to {\it precession-average} the evolution equations for $L$ and $J$ on the radiation-reaction time $t_{\rm RR}$.
These precession-averaged equations give us the ability to efficiently evolve BBH spin directions from formation to near
merger, which is essential to the study of both stellar-mass and supermassive BBHs.  Our previous work
\cite{2013PhRvD..87j4028G,2014PhRvD..89l4025G} revealed that initial spin directions imprinted by the astrophysics of
BBH formation leave detectable GW signatures. The new solutions derived in this {\it Letter} will greatly expand our
capability to explore such formation models.  Supermassive BBH spins will also precess many times before merger
\cite{1980Natur.287..307B}; these solutions will help us predict final-spin distributions for different models of
supermassive black-hole growth \cite{2003ApJ...585L.101H,2008ApJ...684..822B,2012MNRAS.423.2533B,2014ApJ...794..104S,Sijacki:2014yfa} as well as
final-kick distributions that depend sensitively on BBH spin directions at merger
\cite{2007PhRvL..98w1101G,2007PhRvL..98w1102C,2007ApJ...659L...5C,2010ApJ...715.1006K,2012PhRvD..85l4049B}.  Finally, our new solutions may help in the construction and interpretation of GWs from generic double-spin binaries,
 a timely development given the likely first direct detection of GWs later this decade.

\noindent{\em Acknowledgments.~--~}We thank Antoine Klein, Tyson
Littenberg and Daniele Trifir\`o for discussions. D.G. is supported by
the UK STFC and the Isaac Newton Studentship of the University of
Cambridge. R.O'S. is supported by NSF grants PHY-0970074 and
PHY-1307429.  E.B.~is supported by NSF CAREER Grant
PHY-1055103. U.S.~is supported by FP7-PEOPLE-2011-CIG Grant
No. 293412, FP7-PEOPLE-2011-IRSES Grant No.295189, SDSC and TACC
through XSEDE Grant No.~PHY-090003 by the NSF, Finis Terrae through
Grant No.~ICTS-CESGA-249, STFC Consolidator Grant No. ST/L000636/1 and
DiRAC's Cosmos Shared Memory system through BIS Grant No.~ST/J005673/1
and STFC Grant Nos.~ST/H008586/1, ST/K00333X/1.  Figures were
generated using the \texttt{Python}-based \texttt{matplotlib} package
\citep{2007CSE.....9...90H}.

\bibliography{morph}

\begin{thebibliography}{42}%
\makeatletter
\providecommand \@ifxundefined [1]{%
 \@ifx{#1\undefined}
}%
\providecommand \@ifnum [1]{%
 \ifnum #1\expandafter \@firstoftwo
 \else \expandafter \@secondoftwo
 \fi
}%
\providecommand \@ifx [1]{%
 \ifx #1\expandafter \@firstoftwo
 \else \expandafter \@secondoftwo
 \fi
}%
\providecommand \natexlab [1]{#1}%
\providecommand \enquote  [1]{``#1''}%
\providecommand \bibnamefont  [1]{#1}%
\providecommand \bibfnamefont [1]{#1}%
\providecommand \citenamefont [1]{#1}%
\providecommand \href@noop [0]{\@secondoftwo}%
\providecommand \href [0]{\begingroup \@sanitize@url \@href}%
\providecommand \@href[1]{\@@startlink{#1}\@@href}%
\providecommand \@@href[1]{\endgroup#1\@@endlink}%
\providecommand \@sanitize@url [0]{\catcode `\\12\catcode `\$12\catcode
  `\&12\catcode `\#12\catcode `\^12\catcode `\_12\catcode `\%12\relax}%
\providecommand \@@startlink[1]{}%
\providecommand \@@endlink[0]{}%
\providecommand \url  [0]{\begingroup\@sanitize@url \@url }%
\providecommand \@url [1]{\endgroup\@href {#1}{\urlprefix }}%
\providecommand \urlprefix  [0]{URL }%
\providecommand \Eprint [0]{\href }%
\providecommand \doibase [0]{http://dx.doi.org/}%
\providecommand \selectlanguage [0]{\@gobble}%
\providecommand \bibinfo  [0]{\@secondoftwo}%
\providecommand \bibfield  [0]{\@secondoftwo}%
\providecommand \translation [1]{[#1]}%
\providecommand \BibitemOpen [0]{}%
\providecommand \bibitemStop [0]{}%
\providecommand \bibitemNoStop [0]{.\EOS\space}%
\providecommand \EOS [0]{\spacefactor3000\relax}%
\providecommand \BibitemShut  [1]{\csname bibitem#1\endcsname}%
\let\auto@bib@innerbib\@empty
\bibitem [{\citenamefont {{Kerr}}(1963)}]{1963PhRvL..11..237K}%
  \BibitemOpen
  \bibfield  {author} {\bibinfo {author} {\bibfnamefont {R.~P.}\ \bibnamefont
  {{Kerr}}},\ }\href {\doibase 10.1103/PhysRevLett.11.237} {\bibfield
  {journal} {\bibinfo  {journal} {Physical Review Letters}\ }\textbf {\bibinfo
  {volume} {11}},\ \bibinfo {pages} {237} (\bibinfo {year} {1963})}\BibitemShut
  {NoStop}%
\bibitem [{\citenamefont {{Pretorius}}(2005)}]{2005PhRvL..95l1101P}%
  \BibitemOpen
  \bibfield  {author} {\bibinfo {author} {\bibfnamefont {F.}~\bibnamefont
  {{Pretorius}}},\ }\href {\doibase 10.1103/PhysRevLett.95.121101} {\bibfield
  {journal} {\bibinfo  {journal} {Physical Review Letters}\ }\textbf {\bibinfo
  {volume} {95}},\ \bibinfo {eid} {121101} (\bibinfo {year} {2005})},\ \Eprint
  {http://arxiv.org/abs/gr-qc/0507014} {gr-qc/0507014} \BibitemShut {NoStop}%
\bibitem [{\citenamefont {{Campanelli}}\ \emph {et~al.}(2006)\citenamefont
  {{Campanelli}}, \citenamefont {{Lousto}}, \citenamefont {{Marronetti}},\ and\
  \citenamefont {{Zlochower}}}]{2006PhRvL..96k1101C}%
  \BibitemOpen
  \bibfield  {author} {\bibinfo {author} {\bibfnamefont {M.}~\bibnamefont
  {{Campanelli}}}, \bibinfo {author} {\bibfnamefont {C.~O.}\ \bibnamefont
  {{Lousto}}}, \bibinfo {author} {\bibfnamefont {P.}~\bibnamefont
  {{Marronetti}}}, \ and\ \bibinfo {author} {\bibfnamefont {Y.}~\bibnamefont
  {{Zlochower}}},\ }\href {\doibase 10.1103/PhysRevLett.96.111101} {\bibfield
  {journal} {\bibinfo  {journal} {Physical Review Letters}\ }\textbf {\bibinfo
  {volume} {96}},\ \bibinfo {eid} {111101} (\bibinfo {year} {2006})},\ \Eprint
  {http://arxiv.org/abs/gr-qc/0511048} {gr-qc/0511048} \BibitemShut {NoStop}%
\bibitem [{\citenamefont {{Baker}}\ \emph {et~al.}(2006)\citenamefont
  {{Baker}}, \citenamefont {{Centrella}}, \citenamefont {{Choi}}, \citenamefont
  {{Koppitz}},\ and\ \citenamefont {{van Meter}}}]{2006PhRvL..96k1102B}%
  \BibitemOpen
  \bibfield  {author} {\bibinfo {author} {\bibfnamefont {J.~G.}\ \bibnamefont
  {{Baker}}}, \bibinfo {author} {\bibfnamefont {J.}~\bibnamefont
  {{Centrella}}}, \bibinfo {author} {\bibfnamefont {D.-I.}\ \bibnamefont
  {{Choi}}}, \bibinfo {author} {\bibfnamefont {M.}~\bibnamefont {{Koppitz}}}, \
  and\ \bibinfo {author} {\bibfnamefont {J.}~\bibnamefont {{van Meter}}},\
  }\href {\doibase 10.1103/PhysRevLett.96.111102} {\bibfield  {journal}
  {\bibinfo  {journal} {Physical Review Letters}\ }\textbf {\bibinfo {volume}
  {96}},\ \bibinfo {eid} {111102} (\bibinfo {year} {2006})},\ \Eprint
  {http://arxiv.org/abs/gr-qc/0511103} {gr-qc/0511103} \BibitemShut {NoStop}%
\bibitem [{\citenamefont {{Barker}}\ and\ \citenamefont
  {{Oconnell}}(1979)}]{1979GReGr..11..149B}%
  \BibitemOpen
  \bibfield  {author} {\bibinfo {author} {\bibfnamefont {B.~M.}\ \bibnamefont
  {{Barker}}}\ and\ \bibinfo {author} {\bibfnamefont {R.~F.}\ \bibnamefont
  {{Oconnell}}},\ }\href {\doibase 10.1007/BF00756587} {\bibfield  {journal}
  {\bibinfo  {journal} {General Relativity and Gravitation}\ }\textbf {\bibinfo
  {volume} {11}},\ \bibinfo {pages} {149} (\bibinfo {year} {1979})}\BibitemShut
  {NoStop}%
\bibitem [{\citenamefont {{Thorne}}\ and\ \citenamefont
  {{Hartle}}(1985)}]{1985PhRvD..31.1815T}%
  \BibitemOpen
  \bibfield  {author} {\bibinfo {author} {\bibfnamefont {K.~S.}\ \bibnamefont
  {{Thorne}}}\ and\ \bibinfo {author} {\bibfnamefont {J.~B.}\ \bibnamefont
  {{Hartle}}},\ }\href {\doibase 10.1103/PhysRevD.31.1815} {\bibfield
  {journal} {\bibinfo  {journal} {\prd}\ }\textbf {\bibinfo {volume} {31}},\
  \bibinfo {pages} {1815} (\bibinfo {year} {1985})}\BibitemShut {NoStop}%
\bibitem [{\citenamefont {{Apostolatos}}\ \emph {et~al.}(1994)\citenamefont
  {{Apostolatos}}, \citenamefont {{Cutler}}, \citenamefont {{Sussman}},\ and\
  \citenamefont {{Thorne}}}]{1994PhRvD..49.6274A}%
  \BibitemOpen
  \bibfield  {author} {\bibinfo {author} {\bibfnamefont {T.~A.}\ \bibnamefont
  {{Apostolatos}}}, \bibinfo {author} {\bibfnamefont {C.}~\bibnamefont
  {{Cutler}}}, \bibinfo {author} {\bibfnamefont {G.~J.}\ \bibnamefont
  {{Sussman}}}, \ and\ \bibinfo {author} {\bibfnamefont {K.~S.}\ \bibnamefont
  {{Thorne}}},\ }\href {\doibase 10.1103/PhysRevD.49.6274} {\bibfield
  {journal} {\bibinfo  {journal} {\prd}\ }\textbf {\bibinfo {volume} {49}},\
  \bibinfo {pages} {6274} (\bibinfo {year} {1994})}\BibitemShut {NoStop}%
\bibitem [{\citenamefont {{Kidder}}(1995)}]{1995PhRvD..52..821K}%
  \BibitemOpen
  \bibfield  {author} {\bibinfo {author} {\bibfnamefont {L.~E.}\ \bibnamefont
  {{Kidder}}},\ }\href {\doibase 10.1103/PhysRevD.52.821} {\bibfield  {journal}
  {\bibinfo  {journal} {\prd}\ }\textbf {\bibinfo {volume} {52}},\ \bibinfo
  {pages} {821} (\bibinfo {year} {1995})},\ \Eprint
  {http://arxiv.org/abs/gr-qc/9506022} {gr-qc/9506022} \BibitemShut {NoStop}%
\bibitem [{\citenamefont {{Racine}}(2008)}]{2008PhRvD..78d4021R}%
  \BibitemOpen
  \bibfield  {author} {\bibinfo {author} {\bibfnamefont {{\'E}.}~\bibnamefont
  {{Racine}}},\ }\href {\doibase 10.1103/PhysRevD.78.044021} {\bibfield
  {journal} {\bibinfo  {journal} {\prd}\ }\textbf {\bibinfo {volume} {78}},\
  \bibinfo {eid} {044021} (\bibinfo {year} {2008})},\ \Eprint
  {http://arxiv.org/abs/0803.1820} {arXiv:0803.1820 [gr-qc]} \BibitemShut
  {NoStop}%
\bibitem [{\citenamefont {{O'Shaughnessy}}\ \emph {et~al.}(2005)\citenamefont
  {{O'Shaughnessy}}, \citenamefont {{Kaplan}}, \citenamefont {{Kalogera}},\
  and\ \citenamefont {{Belczynski}}}]{2005ApJ...632.1035O}%
  \BibitemOpen
  \bibfield  {author} {\bibinfo {author} {\bibfnamefont {R.}~\bibnamefont
  {{O'Shaughnessy}}}, \bibinfo {author} {\bibfnamefont {J.}~\bibnamefont
  {{Kaplan}}}, \bibinfo {author} {\bibfnamefont {V.}~\bibnamefont
  {{Kalogera}}}, \ and\ \bibinfo {author} {\bibfnamefont {K.}~\bibnamefont
  {{Belczynski}}},\ }\href {\doibase 10.1086/444346} {\bibfield  {journal}
  {\bibinfo  {journal} {\apj}\ }\textbf {\bibinfo {volume} {632}},\ \bibinfo
  {pages} {1035} (\bibinfo {year} {2005})},\ \Eprint
  {http://arxiv.org/abs/astro-ph/0503219} {astro-ph/0503219} \BibitemShut
  {NoStop}%
\bibitem [{\citenamefont {{Belczynski}}\ \emph {et~al.}(2008)\citenamefont
  {{Belczynski}}, \citenamefont {{Taam}}, \citenamefont {{Rantsiou}},\ and\
  \citenamefont {{van der Sluys}}}]{2008ApJ...682..474B}%
  \BibitemOpen
  \bibfield  {author} {\bibinfo {author} {\bibfnamefont {K.}~\bibnamefont
  {{Belczynski}}}, \bibinfo {author} {\bibfnamefont {R.~E.}\ \bibnamefont
  {{Taam}}}, \bibinfo {author} {\bibfnamefont {E.}~\bibnamefont {{Rantsiou}}},
  \ and\ \bibinfo {author} {\bibfnamefont {M.}~\bibnamefont {{van der
  Sluys}}},\ }\href {\doibase 10.1086/589609} {\bibfield  {journal} {\bibinfo
  {journal} {\apj}\ }\textbf {\bibinfo {volume} {682}},\ \bibinfo {pages} {474}
  (\bibinfo {year} {2008})},\ \Eprint {http://arxiv.org/abs/astro-ph/0703131}
  {astro-ph/0703131} \BibitemShut {NoStop}%
\bibitem [{\citenamefont {{Fragos}}\ \emph {et~al.}(2010)\citenamefont
  {{Fragos}}, \citenamefont {{Tremmel}}, \citenamefont {{Rantsiou}},\ and\
  \citenamefont {{Belczynski}}}]{2010ApJ...719L..79F}%
  \BibitemOpen
  \bibfield  {author} {\bibinfo {author} {\bibfnamefont {T.}~\bibnamefont
  {{Fragos}}}, \bibinfo {author} {\bibfnamefont {M.}~\bibnamefont {{Tremmel}}},
  \bibinfo {author} {\bibfnamefont {E.}~\bibnamefont {{Rantsiou}}}, \ and\
  \bibinfo {author} {\bibfnamefont {K.}~\bibnamefont {{Belczynski}}},\ }\href
  {\doibase 10.1088/2041-8205/719/1/L79} {\bibfield  {journal} {\bibinfo
  {journal} {\apjl}\ }\textbf {\bibinfo {volume} {719}},\ \bibinfo {pages}
  {L79} (\bibinfo {year} {2010})},\ \Eprint {http://arxiv.org/abs/1001.1107}
  {arXiv:1001.1107 [astro-ph.HE]} \BibitemShut {NoStop}%
\bibitem [{\citenamefont {{Gerosa}}\ \emph {et~al.}(2013)\citenamefont
  {{Gerosa}}, \citenamefont {{Kesden}}, \citenamefont {{Berti}}, \citenamefont
  {{O'Shaughnessy}},\ and\ \citenamefont {{Sperhake}}}]{2013PhRvD..87j4028G}%
  \BibitemOpen
  \bibfield  {author} {\bibinfo {author} {\bibfnamefont {D.}~\bibnamefont
  {{Gerosa}}}, \bibinfo {author} {\bibfnamefont {M.}~\bibnamefont {{Kesden}}},
  \bibinfo {author} {\bibfnamefont {E.}~\bibnamefont {{Berti}}}, \bibinfo
  {author} {\bibfnamefont {R.}~\bibnamefont {{O'Shaughnessy}}}, \ and\ \bibinfo
  {author} {\bibfnamefont {U.}~\bibnamefont {{Sperhake}}},\ }\href {\doibase
  10.1103/PhysRevD.87.104028} {\bibfield  {journal} {\bibinfo  {journal}
  {\prd}\ }\textbf {\bibinfo {volume} {87}},\ \bibinfo {eid} {104028} (\bibinfo
  {year} {2013})},\ \Eprint {http://arxiv.org/abs/1302.4442} {arXiv:1302.4442
  [gr-qc]} \BibitemShut {NoStop}%
\bibitem [{\citenamefont {{Hughes}}\ and\ \citenamefont
  {{Blandford}}(2003)}]{2003ApJ...585L.101H}%
  \BibitemOpen
  \bibfield  {author} {\bibinfo {author} {\bibfnamefont {S.~A.}\ \bibnamefont
  {{Hughes}}}\ and\ \bibinfo {author} {\bibfnamefont {R.~D.}\ \bibnamefont
  {{Blandford}}},\ }\href {\doibase 10.1086/375495} {\bibfield  {journal}
  {\bibinfo  {journal} {\apjl}\ }\textbf {\bibinfo {volume} {585}},\ \bibinfo
  {pages} {L101} (\bibinfo {year} {2003})},\ \Eprint
  {http://arxiv.org/abs/astro-ph/0208484} {astro-ph/0208484} \BibitemShut
  {NoStop}%
\bibitem [{\citenamefont {{Berti}}\ and\ \citenamefont
  {{Volonteri}}(2008)}]{2008ApJ...684..822B}%
  \BibitemOpen
  \bibfield  {author} {\bibinfo {author} {\bibfnamefont {E.}~\bibnamefont
  {{Berti}}}\ and\ \bibinfo {author} {\bibfnamefont {M.}~\bibnamefont
  {{Volonteri}}},\ }\href {\doibase 10.1086/590379} {\bibfield  {journal}
  {\bibinfo  {journal} {\apj}\ }\textbf {\bibinfo {volume} {684}},\ \bibinfo
  {pages} {822} (\bibinfo {year} {2008})},\ \Eprint
  {http://arxiv.org/abs/0802.0025} {arXiv:0802.0025} \BibitemShut {NoStop}%
\bibitem [{\citenamefont {{Barausse}}(2012)}]{2012MNRAS.423.2533B}%
  \BibitemOpen
  \bibfield  {author} {\bibinfo {author} {\bibfnamefont {E.}~\bibnamefont
  {{Barausse}}},\ }\href {\doibase 10.1111/j.1365-2966.2012.21057.x} {\bibfield
   {journal} {\bibinfo  {journal} {\mnras}\ }\textbf {\bibinfo {volume}
  {423}},\ \bibinfo {pages} {2533} (\bibinfo {year} {2012})},\ \Eprint
  {http://arxiv.org/abs/1201.5888} {arXiv:1201.5888} \BibitemShut {NoStop}%
\bibitem [{\citenamefont {{Sesana}}\ \emph {et~al.}(2014)\citenamefont
  {{Sesana}}, \citenamefont {{Barausse}}, \citenamefont {{Dotti}},\ and\
  \citenamefont {{Rossi}}}]{2014ApJ...794..104S}%
  \BibitemOpen
  \bibfield  {author} {\bibinfo {author} {\bibfnamefont {A.}~\bibnamefont
  {{Sesana}}}, \bibinfo {author} {\bibfnamefont {E.}~\bibnamefont
  {{Barausse}}}, \bibinfo {author} {\bibfnamefont {M.}~\bibnamefont {{Dotti}}},
  \ and\ \bibinfo {author} {\bibfnamefont {E.~M.}\ \bibnamefont {{Rossi}}},\
  }\href {\doibase 10.1088/0004-637X/794/2/104} {\bibfield  {journal} {\bibinfo
   {journal} {\apj}\ }\textbf {\bibinfo {volume} {794}},\ \bibinfo {eid} {104}
  (\bibinfo {year} {2014})},\ \Eprint {http://arxiv.org/abs/1402.7088}
  {arXiv:1402.7088} \BibitemShut {NoStop}%
\bibitem [{\citenamefont {Sijacki}\ \emph {et~al.}(2014)\citenamefont
  {Sijacki}, \citenamefont {Vogelsberger}, \citenamefont {Genel}, \citenamefont
  {Springel}, \citenamefont {Torrey} \emph {et~al.}}]{Sijacki:2014yfa}%
  \BibitemOpen
  \bibfield  {author} {\bibinfo {author} {\bibfnamefont {D.}~\bibnamefont
  {Sijacki}}, \bibinfo {author} {\bibfnamefont {M.}~\bibnamefont
  {Vogelsberger}}, \bibinfo {author} {\bibfnamefont {S.}~\bibnamefont {Genel}},
  \bibinfo {author} {\bibfnamefont {V.}~\bibnamefont {Springel}}, \bibinfo
  {author} {\bibfnamefont {P.}~\bibnamefont {Torrey}},  \emph {et~al.},\
  }\href@noop {} {\  (\bibinfo {year} {2014})},\ \bibinfo {note}
  {arXiv:1408.6842 [astro-ph]}\BibitemShut {NoStop}%
\bibitem [{\citenamefont {{Harry}}\ and\ \citenamefont {{LIGO Scientific
  Collaboration}}(2010)}]{2010CQGra..27h4006H}%
  \BibitemOpen
  \bibfield  {author} {\bibinfo {author} {\bibfnamefont {G.~M.}\ \bibnamefont
  {{Harry}}}\ and\ \bibinfo {author} {\bibnamefont {{LIGO Scientific
  Collaboration}}},\ }\href {\doibase 10.1088/0264-9381/27/8/084006} {\bibfield
   {journal} {\bibinfo  {journal} {Classical and Quantum Gravity}\ }\textbf
  {\bibinfo {volume} {27}},\ \bibinfo {eid} {084006} (\bibinfo {year}
  {2010})}\BibitemShut {NoStop}%
\bibitem [{\citenamefont {{Unnikrishnan}}(2013)}]{2013IJMPD..2241010U}%
  \BibitemOpen
  \bibfield  {author} {\bibinfo {author} {\bibfnamefont {C.~S.}\ \bibnamefont
  {{Unnikrishnan}}},\ }\href {\doibase 10.1142/S0218271813410101} {\bibfield
  {journal} {\bibinfo  {journal} {International Journal of Modern Physics D}\
  }\textbf {\bibinfo {volume} {22}},\ \bibinfo {eid} {1341010} (\bibinfo {year}
  {2013})}\BibitemShut {NoStop}%
\bibitem [{\citenamefont {{Somiya}}(2012)}]{2012CQGra..29l4007S}%
  \BibitemOpen
  \bibfield  {author} {\bibinfo {author} {\bibfnamefont {K.}~\bibnamefont
  {{Somiya}}},\ }\href {\doibase 10.1088/0264-9381/29/12/124007} {\bibfield
  {journal} {\bibinfo  {journal} {Classical and Quantum Gravity}\ }\textbf
  {\bibinfo {volume} {29}},\ \bibinfo {eid} {124007} (\bibinfo {year}
  {2012})},\ \Eprint {http://arxiv.org/abs/1111.7185} {arXiv:1111.7185 [gr-qc]}
  \BibitemShut {NoStop}%
\bibitem [{\citenamefont {{Punturo}}\ \emph {et~al.}(2010)\citenamefont
  {{Punturo}}, \citenamefont {{Abernathy}}, \citenamefont {{Acernese}},
  \citenamefont {{Allen}}, \citenamefont {{Andersson}}, \citenamefont {{Arun}},
  \citenamefont {{Barone}}, \citenamefont {{Barr}}, \citenamefont
  {{Barsuglia}}, \citenamefont {{Beker}} \emph {et~al.}}]{2010CQGra..27s4002P}%
  \BibitemOpen
  \bibfield  {author} {\bibinfo {author} {\bibfnamefont {M.}~\bibnamefont
  {{Punturo}}}, \bibinfo {author} {\bibfnamefont {M.}~\bibnamefont
  {{Abernathy}}}, \bibinfo {author} {\bibfnamefont {F.}~\bibnamefont
  {{Acernese}}}, \bibinfo {author} {\bibfnamefont {B.}~\bibnamefont {{Allen}}},
  \bibinfo {author} {\bibfnamefont {N.}~\bibnamefont {{Andersson}}}, \bibinfo
  {author} {\bibfnamefont {K.}~\bibnamefont {{Arun}}}, \bibinfo {author}
  {\bibfnamefont {F.}~\bibnamefont {{Barone}}}, \bibinfo {author}
  {\bibfnamefont {B.}~\bibnamefont {{Barr}}}, \bibinfo {author} {\bibfnamefont
  {M.}~\bibnamefont {{Barsuglia}}}, \bibinfo {author} {\bibnamefont {{Beker}}},
   \emph {et~al.},\ }\href {\doibase 10.1088/0264-9381/27/19/194002} {\bibfield
   {journal} {\bibinfo  {journal} {Classical and Quantum Gravity}\ }\textbf
  {\bibinfo {volume} {27}},\ \bibinfo {eid} {194002} (\bibinfo {year}
  {2010})}\BibitemShut {NoStop}%
\bibitem [{\citenamefont {{Amaro-Seoane}}\ \emph {et~al.}(2013)\citenamefont
  {{Amaro-Seoane}} \emph {et~al.}}]{2013GWN.....6....4A}%
  \BibitemOpen
  \bibfield  {author} {\bibinfo {author} {\bibfnamefont {P.}~\bibnamefont
  {{Amaro-Seoane}}} \emph {et~al.},\ }\href@noop {} {\bibfield  {journal}
  {\bibinfo  {journal} {GW Notes, Vol.~6, p.~4-110}\ }\textbf {\bibinfo
  {volume} {6}},\ \bibinfo {pages} {4} (\bibinfo {year} {2013})},\ \Eprint
  {http://arxiv.org/abs/1201.3621} {arXiv:1201.3621 [astro-ph.CO]} \BibitemShut
  {NoStop}%
\bibitem [{\citenamefont {{Barausse}}\ \emph {et~al.}(2014)\citenamefont
  {{Barausse}}, \citenamefont {{Bellovary}}, \citenamefont {{Berti}},
  \citenamefont {{Holley-Bockelmann}}, \citenamefont {{Farris}}, \citenamefont
  {{Sathyaprakash}},\ and\ \citenamefont {{Sesana}}}]{2014arXiv1410.2907B}%
  \BibitemOpen
  \bibfield  {author} {\bibinfo {author} {\bibfnamefont {E.}~\bibnamefont
  {{Barausse}}}, \bibinfo {author} {\bibfnamefont {J.}~\bibnamefont
  {{Bellovary}}}, \bibinfo {author} {\bibfnamefont {E.}~\bibnamefont
  {{Berti}}}, \bibinfo {author} {\bibfnamefont {K.}~\bibnamefont
  {{Holley-Bockelmann}}}, \bibinfo {author} {\bibfnamefont {B.}~\bibnamefont
  {{Farris}}}, \bibinfo {author} {\bibfnamefont {B.}~\bibnamefont
  {{Sathyaprakash}}}, \ and\ \bibinfo {author} {\bibfnamefont {A.}~\bibnamefont
  {{Sesana}}},\ }\href@noop {} {\bibfield  {journal} {\bibinfo  {journal}
  {ArXiv e-prints}\ } (\bibinfo {year} {2014})},\ \Eprint
  {http://arxiv.org/abs/1410.2907} {arXiv:1410.2907 [astro-ph.HE]} \BibitemShut
  {NoStop}%
\bibitem [{\citenamefont {{Gerosa}}\ \emph
  {et~al.}(2014{\natexlab{a}})\citenamefont {{Gerosa}}, \citenamefont
  {{O'Shaughnessy}}, \citenamefont {{Kesden}}, \citenamefont {{Berti}},\ and\
  \citenamefont {{Sperhake}}}]{2014PhRvD..89l4025G}%
  \BibitemOpen
  \bibfield  {author} {\bibinfo {author} {\bibfnamefont {D.}~\bibnamefont
  {{Gerosa}}}, \bibinfo {author} {\bibfnamefont {R.}~\bibnamefont
  {{O'Shaughnessy}}}, \bibinfo {author} {\bibfnamefont {M.}~\bibnamefont
  {{Kesden}}}, \bibinfo {author} {\bibfnamefont {E.}~\bibnamefont {{Berti}}}, \
  and\ \bibinfo {author} {\bibfnamefont {U.}~\bibnamefont {{Sperhake}}},\
  }\href {\doibase 10.1103/PhysRevD.89.124025} {\bibfield  {journal} {\bibinfo
  {journal} {\prd}\ }\textbf {\bibinfo {volume} {89}},\ \bibinfo {eid} {124025}
  (\bibinfo {year} {2014}{\natexlab{a}})},\ \Eprint
  {http://arxiv.org/abs/1403.7147} {arXiv:1403.7147 [gr-qc]} \BibitemShut
  {NoStop}%
\bibitem [{\citenamefont {{Peters}}\ and\ \citenamefont
  {{Mathews}}(1963)}]{1963PhRv..131..435P}%
  \BibitemOpen
  \bibfield  {author} {\bibinfo {author} {\bibfnamefont {P.~C.}\ \bibnamefont
  {{Peters}}}\ and\ \bibinfo {author} {\bibfnamefont {J.}~\bibnamefont
  {{Mathews}}},\ }\href {\doibase 10.1103/PhysRev.131.435} {\bibfield
  {journal} {\bibinfo  {journal} {Physical Review}\ }\textbf {\bibinfo {volume}
  {131}},\ \bibinfo {pages} {435} (\bibinfo {year} {1963})}\BibitemShut
  {NoStop}%
\bibitem [{\citenamefont {{Damour}}(2001)}]{2001PhRvD..64l4013D}%
  \BibitemOpen
  \bibfield  {author} {\bibinfo {author} {\bibfnamefont {T.}~\bibnamefont
  {{Damour}}},\ }\href {\doibase 10.1103/PhysRevD.64.124013} {\bibfield
  {journal} {\bibinfo  {journal} {\prd}\ }\textbf {\bibinfo {volume} {64}},\
  \bibinfo {pages} {124013} (\bibinfo {year} {2001})},\ \Eprint
  {http://arxiv.org/abs/gr-qc/0103018} {gr-qc/0103018} \BibitemShut {NoStop}%
\bibitem [{\citenamefont {{Kesden}}\ \emph
  {et~al.}(2010{\natexlab{a}})\citenamefont {{Kesden}}, \citenamefont
  {{Sperhake}},\ and\ \citenamefont {{Berti}}}]{2010PhRvD..81h4054K}%
  \BibitemOpen
  \bibfield  {author} {\bibinfo {author} {\bibfnamefont {M.}~\bibnamefont
  {{Kesden}}}, \bibinfo {author} {\bibfnamefont {U.}~\bibnamefont
  {{Sperhake}}}, \ and\ \bibinfo {author} {\bibfnamefont {E.}~\bibnamefont
  {{Berti}}},\ }\href {\doibase 10.1103/PhysRevD.81.084054} {\bibfield
  {journal} {\bibinfo  {journal} {\prd}\ }\textbf {\bibinfo {volume} {81}},\
  \bibinfo {eid} {084054} (\bibinfo {year} {2010}{\natexlab{a}})},\ \Eprint
  {http://arxiv.org/abs/1002.2643} {arXiv:1002.2643 [astro-ph.GA]} \BibitemShut
  {NoStop}%
\bibitem [{\citenamefont {{Gerosa}}\ \emph
  {et~al.}(2014{\natexlab{b}})\citenamefont {{Gerosa}}, \citenamefont
  {{Kesden}}, \citenamefont {{Berti}}, \citenamefont {{O'Shaughnessy}},\ and\
  \citenamefont {{Sperhake}}}]{GerosaPrep}%
  \BibitemOpen
  \bibfield  {author} {\bibinfo {author} {\bibfnamefont {D.}~\bibnamefont
  {{Gerosa}}}, \bibinfo {author} {\bibfnamefont {M.}~\bibnamefont {{Kesden}}},
  \bibinfo {author} {\bibfnamefont {E.}~\bibnamefont {{Berti}}}, \bibinfo
  {author} {\bibfnamefont {R.}~\bibnamefont {{O'Shaughnessy}}}, \ and\ \bibinfo
  {author} {\bibfnamefont {U.}~\bibnamefont {{Sperhake}}},\ }\href@noop {} {\
  (\bibinfo {year} {2014}{\natexlab{b}})},\ \bibinfo {note} {in
  preparation}\BibitemShut {NoStop}%
\bibitem [{\citenamefont {{Schnittman}}(2004)}]{2004PhRvD..70l4020S}%
  \BibitemOpen
  \bibfield  {author} {\bibinfo {author} {\bibfnamefont {J.~D.}\ \bibnamefont
  {{Schnittman}}},\ }\href {\doibase 10.1103/PhysRevD.70.124020} {\bibfield
  {journal} {\bibinfo  {journal} {\prd}\ }\textbf {\bibinfo {volume} {70}},\
  \bibinfo {eid} {124020} (\bibinfo {year} {2004})},\ \Eprint
  {http://arxiv.org/abs/astro-ph/0409174} {astro-ph/0409174} \BibitemShut
  {NoStop}%
\bibitem [{\citenamefont {{Faye}}\ \emph {et~al.}(2006)\citenamefont {{Faye}},
  \citenamefont {{Blanchet}},\ and\ \citenamefont
  {{Buonanno}}}]{2006PhRvD..74j4033F}%
  \BibitemOpen
  \bibfield  {author} {\bibinfo {author} {\bibfnamefont {G.}~\bibnamefont
  {{Faye}}}, \bibinfo {author} {\bibfnamefont {L.}~\bibnamefont {{Blanchet}}},
  \ and\ \bibinfo {author} {\bibfnamefont {A.}~\bibnamefont {{Buonanno}}},\
  }\href {\doibase 10.1103/PhysRevD.74.104033} {\bibfield  {journal} {\bibinfo
  {journal} {\prd}\ }\textbf {\bibinfo {volume} {74}},\ \bibinfo {eid} {104033}
  (\bibinfo {year} {2006})},\ \Eprint {http://arxiv.org/abs/gr-qc/0605139}
  {gr-qc/0605139} \BibitemShut {NoStop}%
\bibitem [{\citenamefont {{Blanchet}}\ \emph {et~al.}(2006)\citenamefont
  {{Blanchet}}, \citenamefont {{Buonanno}},\ and\ \citenamefont
  {{Faye}}}]{2006PhRvD..74j4034B}%
  \BibitemOpen
  \bibfield  {author} {\bibinfo {author} {\bibfnamefont {L.}~\bibnamefont
  {{Blanchet}}}, \bibinfo {author} {\bibfnamefont {A.}~\bibnamefont
  {{Buonanno}}}, \ and\ \bibinfo {author} {\bibfnamefont {G.}~\bibnamefont
  {{Faye}}},\ }\href {\doibase 10.1103/PhysRevD.74.104034} {\bibfield
  {journal} {\bibinfo  {journal} {\prd}\ }\textbf {\bibinfo {volume} {74}},\
  \bibinfo {eid} {104034} (\bibinfo {year} {2006})},\ \Eprint
  {http://arxiv.org/abs/gr-qc/0605140} {gr-qc/0605140} \BibitemShut {NoStop}%
\bibitem [{\citenamefont {{Buonanno}}\ \emph {et~al.}(2008)\citenamefont
  {{Buonanno}}, \citenamefont {{Kidder}},\ and\ \citenamefont
  {{Lehner}}}]{2008PhRvD..77b6004B}%
  \BibitemOpen
  \bibfield  {author} {\bibinfo {author} {\bibfnamefont {A.}~\bibnamefont
  {{Buonanno}}}, \bibinfo {author} {\bibfnamefont {L.~E.}\ \bibnamefont
  {{Kidder}}}, \ and\ \bibinfo {author} {\bibfnamefont {L.}~\bibnamefont
  {{Lehner}}},\ }\href {\doibase 10.1103/PhysRevD.77.026004} {\bibfield
  {journal} {\bibinfo  {journal} {\prd}\ }\textbf {\bibinfo {volume} {77}},\
  \bibinfo {eid} {026004} (\bibinfo {year} {2008})},\ \Eprint
  {http://arxiv.org/abs/0709.3839} {arXiv:0709.3839} \BibitemShut {NoStop}%
\bibitem [{\citenamefont {{Barausse}}\ and\ \citenamefont
  {{Rezzolla}}(2009)}]{2009ApJ...704L..40B}%
  \BibitemOpen
  \bibfield  {author} {\bibinfo {author} {\bibfnamefont {E.}~\bibnamefont
  {{Barausse}}}\ and\ \bibinfo {author} {\bibfnamefont {L.}~\bibnamefont
  {{Rezzolla}}},\ }\href {\doibase 10.1088/0004-637X/704/1/L40} {\bibfield
  {journal} {\bibinfo  {journal} {\apjl}\ }\textbf {\bibinfo {volume} {704}},\
  \bibinfo {pages} {L40} (\bibinfo {year} {2009})},\ \Eprint
  {http://arxiv.org/abs/0904.2577} {arXiv:0904.2577 [gr-qc]} \BibitemShut
  {NoStop}%
\bibitem [{\citenamefont {{Zhao}}\ and\ \citenamefont
  {{Kesden}}(2014)}]{ZhaoPrep}%
  \BibitemOpen
  \bibfield  {author} {\bibinfo {author} {\bibfnamefont {X.}~\bibnamefont
  {{Zhao}}}\ and\ \bibinfo {author} {\bibfnamefont {M.}~\bibnamefont
  {{Kesden}}},\ }\href@noop {} {\  (\bibinfo {year} {2014})},\ \bibinfo {note}
  {in preparation}\BibitemShut {NoStop}%
\bibitem [{\citenamefont {{Begelman}}\ \emph {et~al.}(1980)\citenamefont
  {{Begelman}}, \citenamefont {{Blandford}},\ and\ \citenamefont
  {{Rees}}}]{1980Natur.287..307B}%
  \BibitemOpen
  \bibfield  {author} {\bibinfo {author} {\bibfnamefont {M.~C.}\ \bibnamefont
  {{Begelman}}}, \bibinfo {author} {\bibfnamefont {R.~D.}\ \bibnamefont
  {{Blandford}}}, \ and\ \bibinfo {author} {\bibfnamefont {M.~J.}\ \bibnamefont
  {{Rees}}},\ }\href {\doibase 10.1038/287307a0} {\bibfield  {journal}
  {\bibinfo  {journal} {\nat}\ }\textbf {\bibinfo {volume} {287}},\ \bibinfo
  {pages} {307} (\bibinfo {year} {1980})}\BibitemShut {NoStop}%
\bibitem [{\citenamefont {{Gonz{\'a}lez}}\ \emph {et~al.}(2007)\citenamefont
  {{Gonz{\'a}lez}}, \citenamefont {{Hannam}}, \citenamefont {{Sperhake}},
  \citenamefont {{Br{\"u}gmann}},\ and\ \citenamefont
  {{Husa}}}]{2007PhRvL..98w1101G}%
  \BibitemOpen
  \bibfield  {author} {\bibinfo {author} {\bibfnamefont {J.~A.}\ \bibnamefont
  {{Gonz{\'a}lez}}}, \bibinfo {author} {\bibfnamefont {M.}~\bibnamefont
  {{Hannam}}}, \bibinfo {author} {\bibfnamefont {U.}~\bibnamefont
  {{Sperhake}}}, \bibinfo {author} {\bibfnamefont {B.}~\bibnamefont
  {{Br{\"u}gmann}}}, \ and\ \bibinfo {author} {\bibfnamefont {S.}~\bibnamefont
  {{Husa}}},\ }\href {\doibase 10.1103/PhysRevLett.98.231101} {\bibfield
  {journal} {\bibinfo  {journal} {Physical Review Letters}\ }\textbf {\bibinfo
  {volume} {98}},\ \bibinfo {eid} {231101} (\bibinfo {year} {2007})},\ \Eprint
  {http://arxiv.org/abs/gr-qc/0702052} {gr-qc/0702052} \BibitemShut {NoStop}%
\bibitem [{\citenamefont {{Campanelli}}\ \emph
  {et~al.}(2007{\natexlab{a}})\citenamefont {{Campanelli}}, \citenamefont
  {{Lousto}}, \citenamefont {{Zlochower}},\ and\ \citenamefont
  {{Merritt}}}]{2007PhRvL..98w1102C}%
  \BibitemOpen
  \bibfield  {author} {\bibinfo {author} {\bibfnamefont {M.}~\bibnamefont
  {{Campanelli}}}, \bibinfo {author} {\bibfnamefont {C.~O.}\ \bibnamefont
  {{Lousto}}}, \bibinfo {author} {\bibfnamefont {Y.}~\bibnamefont
  {{Zlochower}}}, \ and\ \bibinfo {author} {\bibfnamefont {D.}~\bibnamefont
  {{Merritt}}},\ }\href {\doibase 10.1103/PhysRevLett.98.231102} {\bibfield
  {journal} {\bibinfo  {journal} {Physical Review Letters}\ }\textbf {\bibinfo
  {volume} {98}},\ \bibinfo {eid} {231102} (\bibinfo {year}
  {2007}{\natexlab{a}})},\ \Eprint {http://arxiv.org/abs/gr-qc/0702133}
  {gr-qc/0702133} \BibitemShut {NoStop}%
\bibitem [{\citenamefont {{Campanelli}}\ \emph
  {et~al.}(2007{\natexlab{b}})\citenamefont {{Campanelli}}, \citenamefont
  {{Lousto}}, \citenamefont {{Zlochower}},\ and\ \citenamefont
  {{Merritt}}}]{2007ApJ...659L...5C}%
  \BibitemOpen
  \bibfield  {author} {\bibinfo {author} {\bibfnamefont {M.}~\bibnamefont
  {{Campanelli}}}, \bibinfo {author} {\bibfnamefont {C.}~\bibnamefont
  {{Lousto}}}, \bibinfo {author} {\bibfnamefont {Y.}~\bibnamefont
  {{Zlochower}}}, \ and\ \bibinfo {author} {\bibfnamefont {D.}~\bibnamefont
  {{Merritt}}},\ }\href {\doibase 10.1086/516712} {\bibfield  {journal}
  {\bibinfo  {journal} {\apjl}\ }\textbf {\bibinfo {volume} {659}},\ \bibinfo
  {pages} {L5} (\bibinfo {year} {2007}{\natexlab{b}})},\ \Eprint
  {http://arxiv.org/abs/gr-qc/0701164} {gr-qc/0701164} \BibitemShut {NoStop}%
\bibitem [{\citenamefont {{Kesden}}\ \emph
  {et~al.}(2010{\natexlab{b}})\citenamefont {{Kesden}}, \citenamefont
  {{Sperhake}},\ and\ \citenamefont {{Berti}}}]{2010ApJ...715.1006K}%
  \BibitemOpen
  \bibfield  {author} {\bibinfo {author} {\bibfnamefont {M.}~\bibnamefont
  {{Kesden}}}, \bibinfo {author} {\bibfnamefont {U.}~\bibnamefont
  {{Sperhake}}}, \ and\ \bibinfo {author} {\bibfnamefont {E.}~\bibnamefont
  {{Berti}}},\ }\href {\doibase 10.1088/0004-637X/715/2/1006} {\bibfield
  {journal} {\bibinfo  {journal} {\apj}\ }\textbf {\bibinfo {volume} {715}},\
  \bibinfo {pages} {1006} (\bibinfo {year} {2010}{\natexlab{b}})},\ \Eprint
  {http://arxiv.org/abs/1003.4993} {arXiv:1003.4993 [astro-ph.CO]} \BibitemShut
  {NoStop}%
\bibitem [{\citenamefont {{Berti}}\ \emph {et~al.}(2012)\citenamefont
  {{Berti}}, \citenamefont {{Kesden}},\ and\ \citenamefont
  {{Sperhake}}}]{2012PhRvD..85l4049B}%
  \BibitemOpen
  \bibfield  {author} {\bibinfo {author} {\bibfnamefont {E.}~\bibnamefont
  {{Berti}}}, \bibinfo {author} {\bibfnamefont {M.}~\bibnamefont {{Kesden}}}, \
  and\ \bibinfo {author} {\bibfnamefont {U.}~\bibnamefont {{Sperhake}}},\
  }\href {\doibase 10.1103/PhysRevD.85.124049} {\bibfield  {journal} {\bibinfo
  {journal} {\prd}\ }\textbf {\bibinfo {volume} {85}},\ \bibinfo {eid} {124049}
  (\bibinfo {year} {2012})},\ \Eprint {http://arxiv.org/abs/1203.2920}
  {arXiv:1203.2920 [astro-ph.HE]} \BibitemShut {NoStop}%
\bibitem [{\citenamefont {{Hunter}}(2007)}]{2007CSE.....9...90H}%
  \BibitemOpen
  \bibfield  {author} {\bibinfo {author} {\bibfnamefont {J.~D.}\ \bibnamefont
  {{Hunter}}},\ }\href {\doibase 10.1109/MCSE.2007.55} {\bibfield  {journal}
  {\bibinfo  {journal} {Computing in Science and Engineering}\ }\textbf
  {\bibinfo {volume} {9}},\ \bibinfo {pages} {90} (\bibinfo {year}
  {2007})}\BibitemShut {NoStop}%
\end{thebibliography}%
\end{document}